\documentclass[12pt]{revtex4}

\usepackage{amsmath}
\usepackage{rotating}
\usepackage{graphicx}
\usepackage{lscape}
\usepackage{amsfonts}
\usepackage{amssymb}
\usepackage{amsthm}
\usepackage[T1]{fontenc}
\usepackage[spanish]{babel}
\usepackage[active]{srcltx}
\usepackage{graphicx}
\usepackage{epsfig}
\usepackage{pb-diagram}
\usepackage{tensor}

\begin{document}
\newcommand{\nl}{{\cal N}}
\newcommand{\cl}{\Phi}
\def\binom#1#2{{#1\choose #2}}
\newcommand{\sas}{ $\hat{S}^{2}$ and $\hat{S}_{z}$ }
\newcommand{\hu}{{\cal A}}
\newcommand{\dm}{{\cal D}}
\newcommand{\idn}{{\cal I}}
\newcommand{\lo}{\Lambda\Omega}




\newtheorem{corollary}{Corollary}
\newtheorem{lemma}{Lemma}
\newtheorem{proposition}{Proposition}
\newtheorem{theorem}{Theorem}


\newenvironment{dem}[1][Proof]{\begin{proof}[{\bf #1}]}{\end{proof}}


\theoremstyle{definition}

\newtheorem{axiom}{Axioma}[section]
\newtheorem{definition}{Definición}[section]
\newtheorem{example}{Ejemplo}[section]
\newtheorem{remark}{Observación}[section]
\newtheorem{exercise}{Ejercicio}[section]
{\swapnumbers\newtheorem{exercisestar}[exercise]{(*)}}


\newcommand{\C}{\ensuremath{\mathbb{C}}}
\newcommand{\N}{\ensuremath{\mathbb{N}}}
\newcommand{\Q}{\ensuremath{\mathbb{Q}}}
\newcommand{\R}{\ensuremath{\mathbb{R}}}
\newcommand{\T}{\ensuremath{\mathbb{T}}}
\newcommand{\Z}{\ensuremath{\mathbb{Z}}}


\newcommand{\abs}[1]{\left\vert #1\right\vert}
\newcommand{\bra}[1]{\left\langle #1\right\vert}
\renewcommand{\dim}[1]{\mathrm{dim}\left( #1\right)}
\newcommand{\set}[1]{\left\{ #1\right\}}
\renewcommand{\ker}[1]{\mathrm{Ker}\pa{#1}}
\newcommand{\ket}[1]{\left\vert #1\right\rangle}
\newcommand{\norm}[1]{\left\|#1\right\|}
\newcommand{\pa}[1]{\left(#1\right)}
\newcommand{\pro}[2]{\left\langle#1|#2\right\rangle}
\newcommand{\proo}[3]{\left\langle#1\left|#2\right|#3\right\rangle}
\newcommand{\ran}[1]{\mathrm{Ran}\pa{#1}}
\newcommand{\tr}[1]{\mathop{\mathrm{Tr}}\pa{#1}}

\newcommand{\Rn}[1][\mathcal{N}]{\mathbb{R}^{#1}}
\newcommand{\Cn}[1][\mathcal{N}]{\C^{#1}}
\newcommand{\Sph}[1][\mathcal{N}]{\mathrm{S}^{#1}}

\newcommand{\codim}[1]{\mathrm{codim}\left( #1\right)}
\newcommand{\diag}{\mathop{\mathrm{diag}}}
\newcommand{\diam}{\mathop{\mathrm{diam}}}
\newcommand{\id}{\mathop{\mathrm{id}}}
\newcommand{\inte}{\mathop{\mathrm{int}}}
\newcommand{\inc}{\mathop{\mathrm{\iota}}}
\newcommand{\sop}{\mathop{\mathrm{sop}}}

\newcommand{\GL}[2][R]{\mathrm{GL}\pa{\mathbb{#1},#2}}
\newcommand{\GLp}[2][R]{\mathrm{GL}_{+}\pa{\mathbb{#1},#2}}
\newcommand{\SL}[2][R]{\mathrm{SL}\pa{\mathbb{#1},#2}}
\newcommand{\Or}[1][\mathcal{N}]{\mathrm{O}\pa{#1}}
\newcommand{\SO}[1][\mathcal{N}]{\mathrm{SO}\pa{#1}}
\newcommand{\Un}[1][\mathcal{N}]{\mathrm{U}\pa{#1}}
\newcommand{\SU}[1][\mathcal{N}]{\mathrm{SU}\pa{#1}}
\newcommand{\Up}[2][R]{\mathrm{Up}\pa{\mathbb{#1},#2}}
\newcommand{\Upp}[2][R]{\mathrm{Up}_{+}\pa{\mathbb{#1},#2}}
\newcommand{\her}[1]{\mathrm{her}\pa{#1}}
\newcommand{\hil}{\mathsf{H}}

\renewcommand{\Re}[1]{\mathrm{Re}\left( #1\right)}
\renewcommand{\Im}[1]{\mathrm{Im}\left( #1\right)}



\large \noindent {\rm {\bf  Chemical descriptors, convexity and
structure of density matrices in molecular systems}}

\vskip 5mm \normalsize \noindent Roberto C. Bochicchio$^{*}$

\vskip 2mm

\noindent {\it Departamento de F\'{\i}sica, Facultad de Ciencias
Exactas y Naturales, Universidad de Buenos Aires and IFIBA, CONICET,
Ciudad Universitaria, 1428, Buenos Aires, Argentina}

\vskip 3mm

\noindent The electron energy and density matrices in molecular
systems are convex in respect of the number of particles. So that,
the chemical descriptors based on their derivatives present the
hamper of discontinuities for isolated systems and consequently
higher order derivatives are undefined. The introduction of the
interaction between the physical domain with an environment induces
a coherent structure for the density matrix in the grand-canonical
formulation suppressing the discontinuities leading to the proper
definitions of the descriptors.



\vskip 60 mm {\small \noindent
\rule{60mm}{0.4mm} \\
\noindent $^{*}$ E-mail address: rboc@df.uba.ar (R. C. Bochicchio)}

\newpage

\normalsize

The electron distribution rearrangements in molecular systems under
the influence of external perturbations, internal conversions,
conformational changes or reactive interactions are closely related
to chemical reactivity which is relevant to understand the molecular
structure.$^{1,2}$ The chemical units involved in these processes
are interacting moieties, atoms or functional groups which undergo a
flux of charge by electron exchange between them.$^{2,3}$ This fact
induces such domains to possess a non-integer (or fractional) number
of electrons which may be interpreted as an ensemble average in a
quantum state of an open system at equilibrium after the
rearrangement process.$^{3,4}$ The description of this phenomena is
performed by means of the fundamental magnitudes, energy, electron
density and their derivatives with respect of the number of
particles $\mathcal{N}$ which incorporates this change at the very
basis of the descriptor definitions.$^{1,3,4}$ It imposes the
knowledge of the energy $\mathcal{E}_{0}^{\mathcal{N}}$ dependence
with this number, where $\mathcal{N}=N \pm \nu$ with $N\in\N$ and
$\nu\in\R$ the electron transferred fraction which lies in the
$\nu\in\pa{0,1}$ interval ($\N$ and $\R$ fields of positive integer
and real numbers, respectively), i.e., their dependence not only at
integer numbers but for all values of $\mathcal{N}$.$^{3,4}$
Nevertheless the concept of reactivity is related to the interaction
between the entities involved in the rearrangements, the quantities
used to describe the behavior of reactive phenomena are usually
evaluated by finite differences with respect to integer number of
the particles of isolated species.$^{1,2}$ This approach ignores the
values for the magnitudes at fractional numbers and consequently the
electron exchange as the onset of chemical behavior. To avoid such
inconsistences, exhaustive works supported by the above mentioned
energy dependence has been performed about essential treatment for
these chemical descriptors from state function approach.$^{5}$ The
most general way to describe a quantum state of a system is by means
of the density matrix (DM) formalism$^{6-8}$ which contains the
complete information about the system, i.e., from which all
properties may be determined.$^{6-8}$ In the case of a system with a
fractional number of particles, the state can not be described by a
pure state neither a canonical ensemble but by an statistical
ensemble of pure states with different number of electrons, i.e., a
grand-canonical like ensemble (GC).$^{3,4}$ For ground states the
dependence of the energy $\mathcal{E}_{0}^{\mathcal{N}}$ and the DMs
is a piecewise-continuous linear functions of $\mathcal{N}$ and only
the bordering integers $N$ and $N \pm 1$ enter in this
ensemble.$^{4}$ Consequently, ground state properties then have the
same dependence (cf. Fig. 1a of Ref. 3) and the first derivatives of
the energy and the density are staircase functions of $\mathcal{N}$,
undefined at the integers and constant in between (cf. Fig. 1b of
Ref. 3). Thus, second derivatives vanish in between and are not
defined for integers.$^{2}$ This behavior has deep consequences, for
instance, electronegativity equalization principle does not apply
and the electronic principles based on hardness has no rigorous
foundations.$^{1-3}$ As mentioned above, attempts to avoid the
inconsistencies has been proposed for pure state formulations$^{5}$
but no a general statistical formulation has been reported
considering the system interaction with the environment, i.e., other
subsystem or reservoir (S-R) interactions. The main objective of
this report is to introduce the rigorous scheme for the DMs
structure$^{3,4}$ within the interaction between the subsystems to
highlight the essence of the reactivity descriptors.





The weighted sum of the complete set of all accesible $M$-electron
pure state density matrices$^{6-8}$
$^{M}D_{\cl^{N}_{k}}=|\cl^{M}_{k}><\cl^{M}_{k}|$ in the mixture,
with $|{\cl^{M}_{k}}>$ the $k-th$ quantum state function in the
antisymmetric $M$-electron Hilbert space ${\cal F}_{M}$,$\; ^{7,8}$
represents the state of the system by the density matrix $D$, whose
carrier space is the Fock space ${\cal
F}=\bigoplus_{M=0}^{\infty}{\cal F}_{M}$ where $\bigoplus$ symbol
indicates direct sum$^{7,8}$, and reads

\begin{equation}
D\; =\; \sum_{M \ge 0} \sum_{k \ge 0}\; \omega_{\cl^{M}_{k}}\;
|\cl^{M}_{k}><\cl^{M}_{k}|; \hspace{0.7cm} \sum_{M \ge 0} \sum_{k
\ge 0}\; \omega_{\cl^{M}_{k}}\;=\; 1; \hspace{0.5cm}
\omega_{\cl^{M}_{k}}\; \ge 0 \label{1}
\end{equation}
\vskip 5mm

\noindent where $\{\omega_{\cl^{M}_{k}}, M \ge 0; k \ge 0\}$ stands
for the set of statistical weights, i.e., probability of occurrence
of the pure state $|\cl^{M}_{k}>$ in the mixture. This state admits
particle number fluctuation and hence the system may posses a
non-integer number of particles. We will refer it as the GC
distribution. $D$ is an Hermitian, positive semi-definite (its
eigenvalues are nonnegative or null), bounded (its elements are
bounded) and finite trace (sum of diagonal elements) matrix. Thus,
because of its probabilistic interpretation it may be normalized to
unity, i.e., $tr(D)=1$.$^{6-8}$

The fundamental chemical concepts are the summary of the physical
information contained in $D$ and described by two types of
quantities called descriptors. On one side are those coming from the
direct integration of a function of the density related to the
classical concepts of chemistry$^{9}$ and on the other side, those
related to the energy/density derivatives of first and higher
orders.$^{1}$ Let us remark at this point that the physical domains
within the molecular structure which possess a fractional number of
particles ${\cal{N}}=N \pm \nu$ described by GC states (Eq. (1)) may
be interpreted as an average.$^{3,4}$ The convex structure of $D$
for ground states may be expressed in two branches each one as a two
state model of $N$ and $N \pm 1$ Hilbert spaces as $D\; =\;
\pa{1-\nu}\; {}^{N}\!{D}_{0}+\nu \; {}^{N \pm 1}\!\!{D}_{0}$$\;
^{3,4}$ with $^{N}\!{D}_{0}=|{\cl^{N}_{0}}><{\cl^{N}_{0}}|$ and
${}^{N \pm 1}\!\!{D}_{0}=|{\cl^{N \pm 1}_{0}}><\cl^{N \pm 1}_{0}|$
the corresponding ground state DMs. The energy is
$\mathcal{E}_{0}^{N\pm\nu}\;=\; \pa{1-\nu}\;
\mathcal{E}_{0}^{N}\;+\; \nu\; \mathcal{E}_{0}^{N+1}$.$\; ^{3,4}$
The onset of the above mentioned inconsistencies are related to the
lack of information in the description about the interaction of the
system (subsystem) with the environment. To avoid it we introduce a
driven interaction potential $\bf U_{\nu}$ which has its origin in
the subsystem fragment within the Atoms in Molecules (AIM)
framework$^{5}$ or reservoir interactions effects$^{6}$ describing
the influence of the environment. The subscript in  $\bf U_{\nu}$
indicates its electron fraction dependence. So that, the Hamiltonian
for each of the two branches, ${\bf H}={\bf H}_{0}+{\bf U}_{\nu}$ in
matrix form is $\mathcal{H}=\;
\mathcal{E}_{0}^{N}|\cl^{N}_{0}><\cl^{N}_{0}|\; +\;
\mathcal{E}_{0}^{N \pm 1}|\cl^{N \pm 1}_{0}><\cl^{N \pm 1}_{0}|\;
+\; \mathcal{U}_{\nu}^{\pm}|\cl^{N}_{0}><\cl^{N \pm 1}_{0}|\; +\;
{\mathcal{U}_{\nu}^{\pm}}^{*}|\cl^{N \pm 1}_{0}><\cl^{N}_{0}|$ where
${\bf H}_{0}$ stands for the isolated system Hamiltonian whose
states produce discontinuities in the high order descriptors.$^{2}$
The interaction potential induces a new solution $\tilde{D}$ which
at equilibrium exchanges electrons at a time independent rate, i.e.,
the electron exchange is constant in time, and hence the solution
may exhibit a coherent structure$^{6}$ expressed as

\begin{equation}
\tilde{D}\; =\; D\; +\; \Delta_{\nu}^{\pm}\; |\cl^{N}_{0}><\cl^{N
\pm 1}_{0}|\; +\; {\Delta_{\nu}^{\pm}}^{*}\; |\cl^{N \pm
1}_{0}><\cl^{N}_{0}| \label{2}
\end{equation}
\vskip 5mm

\noindent where the first term of the r.h.s. represents the solution
for ${\bf H}_{0}$,$^{4}$ while the last two terms correspond to the
coupling interaction of the $|\cl^{N}_{0}>$ and $|\cl^{N \pm
1}_{0}>$ states. Within this scenario, the energy of the system
under the influence of the environment results

\begin{equation}
\tilde{\mathcal{E}}_{0}^{N\pm\nu}\;=\; Tr(\mathcal{H}\tilde{D})\;
=\; \mathcal{E}_{0}^{N\pm\nu}\; +\; 2
\mathbb{R}e({\mathcal{U}_{\nu}^{\pm}}^{*}{\Delta_{\nu}^{\pm}})
\label{3}
\end{equation}
\vskip 5mm

\noindent where the symbol $\mathbb{R}e$ stands for the real part of
the complex number
${\mathcal{U}_{\nu}^{\pm}}^{*}\Delta_{\nu}^{\pm}$. Therefore, this
term may introduce a $\nu$-nonlinearity dependence for the energy
and the DM and thus enable us to perform the calculation of the
chemical descriptor of arbitrary order avoiding the discontinuity
problem. To be more concrete let us write two important descriptors
to show these ideas. The chemical potential defined by$^{1-4}$

\begin{equation}
\tilde{\mu}^{\pm}=\; \left(\frac{\partial \tilde{\mathcal{E}}_{0}^{
\mathcal{N}}}{\partial \mathcal{N}} \right)_{v} =\; \pm
\left(\frac{\partial \tilde{\mathcal{E}}_{0}^{N\pm\nu}}{\partial
\nu} \right)_{v} \label{4}
\end{equation}
\vskip 5mm \noindent at constant $v$ (external field), becomes

\begin{equation}
\tilde{\mu}^{\pm}=\; {\mu}^{\pm} \pm \; 2 \left( \frac{\partial
\mathbb{R}e({\mathcal{U}_{\nu}^{\pm}}^{*}{\Delta_{\nu}^{\pm}})}
{\partial \nu} \right)_{v} \label{5}
\end{equation}
\vskip 5mm \noindent where ${\mu}^{\pm}$ stands for the
${\mu}^{+}=-EA$ and ${\mu}^{-}=-IP$, electron affinity and
ionization potential, respectively and the second term in Eq. (5)
shows the interaction contribution to the chemical potential which
avoids the discontinuity$^{2}$ and admits the equalization principle
to be fulfilled$^{1}$. For that goal, let us consider two fragments
$\Omega_{A}$ and $\Omega_{B}$ within a molecular framework which at
equilibrium must obey the condition
$\tilde{\mu}^{+}_{\Omega_{A}}=\tilde{\mu}^{-}_{\Omega_{B}}$, i.e.,
the chemical potential of the donor fragment must be equal to that
of the acceptor fragment and it is the second term of the r.h.s. of
Eq. (5) which enable this condition. The hardness$^{1}$ which would
vanish because of the chemical potential discontinuity for an
isolated system, i.e., without interaction with an
environment$^{2}$, becomes non null due to the S-R interaction and
reads

\begin{equation}
\tilde{\eta}=\; \frac{1}{2}\left(\frac{\partial^{2}
\tilde{\mathcal{E}}_{0}^{ \mathcal{N}}}{\partial^{2} \mathcal{N}}
\right)_{v} =\; \pm  \left( \frac{\partial^{2}
\mathbb{R}e({\mathcal{U}_{\nu}^{\pm}}^{*}{\Delta_{\nu}^{\pm}})}
{\partial^{2} \nu} \right)_{v} \label{6}
\end{equation}
\vskip 5mm \noindent

\noindent Let us finally to mention some concluding remarks. This
approach is general regarding the DMs may be calculated from any
arbitrary methodology and it depends only of the model used for
describing the interaction with the environment; also this
formulation permits to recover the piece-wise dependence as the
interaction vanishes, i.e., $\bf U_{\nu} \rightarrow 0$.

\vskip 3mm

\noindent {\bf Acknowledgments} Financially supported by Projects
20020130100226BA (UBA, Argentina) and 11220090100061 (CONICET,
Argentina)

\vskip 5mm

{\small

\noindent $^{1}$ P. Geerlings, F. De Proft and W. Langenaeker, Chem.
Rev. {\bf 103}, 1793 (2003).

\noindent $^{2}$ M. H. Cohen and A. Wasserman, J. Phys. Chem. A {\bf
111}, 2229 (2007) and references.

\noindent $^{3}$ J. P. Perdew, R. G. Parr, M. Levy and J. Balduz,
Jr., Phys. Rev. Lett. {\bf 49}, 1691 (1982).

\noindent $^{4}$ R. C. Bochicchio, R. A. Miranda-Quintana and D.
Rial, J. Chem. Phys. {\bf 139}, 191101 (2013) and references.

\noindent $^{5}$ S. M. Valone, J. Phys. Chem. Lett. {\bf 2}, 2618
(2011) and references.

\noindent $^{6}$ K. Blum, {\it Density Matrix Theory and
Applications} (Plenum, New York, 1981).

\noindent $^{7}$ A. J. Coleman and V. I. Yukalov, {\it Reduced
Density Matrices: Coulson's Challenge} (Springer, New York, 2000).

\noindent $^{8}$ {\it Reduced-Density-Matrix Mechanics with
Applications to Many-Electron Atoms and Molecules, Advances in
Chemical Physics} Vol. 134, edited by D. A. Mazziotti
(Wiley-Intersience, Hoboken, 2007)

\noindent $^{9}$ R. C. Bochicchio, L. Lain and A. Torre, Chem. Phys.
Lett., {\bf 375}, 45 (2003) and references.

\end{document}